\documentclass[american,aps,preprint,prb,showpacs]{revtex4}
\usepackage[T1]{fontenc}
\usepackage[latin9]{inputenc}
\usepackage{verbatim}
\usepackage{amsmath}
\usepackage{graphicx}
\usepackage{esint}

\makeatletter

\providecommand{\tabularnewline}{\\}

\@ifundefined{textcolor}{}
{%
 \definecolor{BLACK}{gray}{0}
 \definecolor{WHITE}{gray}{1}
 \definecolor{RED}{rgb}{1,0,0}
 \definecolor{GREEN}{rgb}{0,1,0}
 \definecolor{BLUE}{rgb}{0,0,1}
 \definecolor{CYAN}{cmyk}{1,0,0,0}
 \definecolor{MAGENTA}{cmyk}{0,1,0,0}
 \definecolor{YELLOW}{cmyk}{0,0,1,0}
 }

\makeatother
\usepackage{dcolumn}

\makeatother

\usepackage{babel}

\begin{document}

\title{Simple Vortex States in Films of Type-I Ginzburg-Landau Superconductor}

\date{\today}

\author{Mark C. Sweeney}

\author{Martin P. Gelfand}

\email{martin.gelfand@colostate.edu}

\affiliation{Department of Physics, Colorado State University, Fort Collins, Colorado
80523--1875 USA}
\begin{abstract}
Sufficiently thin films of type-I superconductor in a perpendicular
magnetic field exhibit a triangular vortex lattice, while thick films
develop an intermediate state. To elucidate what happens between these
two regimes, precise numerical calculations have been made within
Ginzburg-Landau theory at $\kappa=0.5$ and 0.25 for a variety of
vortex lattice structures with one flux quantum per unit cell. The
phase diagram in the space of mean induction and film thickness includes
a narrow wedge in which a square lattice is stable, surrounded by
the domain of stability of the triangular lattice at thinner films/lower
fields and, on the other side, rectangular lattices with continuously
varying aspect ratio. The vortex lattice has an anomalously small
shear modulus within and close to the square lattice phase.
\end{abstract}


\maketitle

\section{Introduction\label{sec:Introduction}}

Thin films of a bulk type-I superconductor subject to a perpendicular
magnetic field can behave like bulk type-II superconductors, in that
they develop a vortex lattice. The pioneering theoretical treatments
in the area, by Tinkham\cite{tinkham_effect_1963} and Maki\cite{maki_fluxoid_1965}
applied Ginzburg-Landau (GL) theory in the vicinity of the critical
field where the order parameter vanishes. Pearl's\cite{pearl_current_1964,Pearl_thesis}
treatment of isolated vortices within London theory, appropriate in
the low-field limit, shows that vortices in a sufficiently thin film
have a long-range repulsion. This repulsion should lead to the development
of a triangular vortex lattice at small fields, and it was shown by
Lasher\cite{lasher_mixed_1967} that a triangular vortex lattice is
also favored near the upper critical field for sufficiently thin films.

On the other hand, sufficiently thick films of type-I superconductors
exhibit the intermediate state. How does the magnetic flux structure
evolve from triangular vortex lattice to intermediate state with increasing
film thickness? In other words, what is the equilibrium phase diagram
for magnetic flux structures, as a function of film thickness and
magnetic field? In real superconductors the picture will necessarily
be complicated by disorder and anisotropy, but the question is interesting
enough within GL theory. Lasher\cite{lasher_mixed_1967} was the first
to address it. He showed that, within linearized GL theory, between
the triangular vortex lattice and the intermediate state there were
a large number of distinct vortex phases, with the triangular lattice
first replaced by a square vortex lattice. Some years later Callaway\cite{callaway_magnetic_1992}
pointed out that Lasher had not considered the most general Abrikosov-type
solutions to the linearized first GL equation, and he carried out
a comprehensive analysis of periodic vortex arrays, valid close to
the upper critical field.

While theoretical analyses of the magnetic flux structure phase diagram
have been restricted to the highest possible fields, interesting experimental
results have appeared at very low fields. Hasegawa \textit{et al.}\cite{hasegawa_magnetic-flux_1991}
applied electron holography to examine the magnetic field in the space
above flux structures in Pb films. They found evidence for vortices
with more than one flux quantum (which they denoted MQF-A) as well
as flux structures that seemed more likely to be associated with normal
regions of finite cross-section (which they denoted MQF-B). {}``Multiply-quantized''
(also known as {}``giant'') vortices are known to arise in various
circumstances. Holes in a superconductor parallel to the field trap
vortices with greater fluxoid number as their radii increase;\cite{Mkrtchyan}
the repulsion of vortices from a film edge can lead to the formation
of a equilibrium giant vortex in the center of a small, thin disk;\cite{Schweigert_Peeters_Deo_98}
metastable giant vortices develop in field-cooling of small cylinders.\cite{Moshchalkov_97}
None of these seem relevant to the experiment of Hasegawa \textit{et
al.}, and the search for stable lattices of multiply-quantized vortex
in the phase diagram for type-I films was one of the motivations for
the present work.

The magnetic flux structure phase diagram is also of fundamental interest,
because vortices in a film with GL parameter $\kappa<1/\sqrt{2}$
comprise an archetypical system with competing interactions. It is
not obvious whether the vortex structures found by Lasher and Callaway
at intermediate thicknesses survive on reducing the magnetic field,
and attempts to experimentally observe such structures would benefit
from theoretical guidance.

It is also noteworthy that a bulk GL superconductor with $\kappa=1/\sqrt{2}$
and at the critical field exhibits massive (in fact complete) degeneracy
with respect to vortex configurations.\cite{Bogomolnyi1} Luk'yanchuk\cite{lukyanchuk_theory_2001}
has carried out a thorough analysis of corrections to the GL functional,
together with deviations of $\kappa$ and the magnetic field from
their critical values, in breaking the degeneracy. He noted that demagnetization
effects also break the degeneracy, but did no calculations along those
lines. A film geometry corresponds to maximum demagnetization, so
it may be interesting to compare the vortex phase diagram for films
with $\kappa\approx1/\sqrt{2}$ with the phase diagrams that follow
from the analysis by Luk'yanchuk.

In this paper we take the first steps towards filling out the magnetic
flux structure phase diagram for the minimal model, isotropic Ginzburg-Landau
theory, of thin film type-I superconductors. The competition between
various phases is delicate at the upper critical field, so accurate
free-energy calculations for different flux structures are necessary.
Consequently, we have followed the approach pioneered by Brandt for
vortex lattices in bulk\cite{brandt_ginsburg-landau_1972} and, more
recently, thin film\cite{brandt_ginzburg-landau_2005} GL superconductors.
The squared magnitude of the order parameter, the supervelocity, and
the magnetic field are represented as linear combinations of appropriate
basis functions. The GL equations then become a set of nonlinear equations
for the coefficients, which are solved by iteration.

Sec.~\ref{sec:Computational-Method} describes the computational
method in more detail. Brandt's papers are quite explicit, so we may
be brief, and highlight the modest changes we made to Brandt's algorithm
for thin films superconductors. The present calculations are restricted
to magnetic flux structures consisting of singly-quantized vortices
in periodic structures with one vortex per unit cell. We have found
that the functional form chosen for the magnetic field in Ref.~\onlinecite{brandt_ginzburg-landau_2005}
limits the accuracy of the magnetic field and consequently the free
energy, and we offer a correct and computationally convenient alternative.
Sec.~\ref{sec:Results} presents the principal results of the calculations,
which are based on evaluating the free energy for a large number of
points in the space of vortex lattices structures, film thickness,
and applied magnetic fields (or, equivalently, mean inductions). Phase
diagrams, free energy densities, and vortex lattice shear moduli are
given for $\kappa=0.5$ and 0.25. Other values of the GL parameter
could have been considered but the calculations become significantly
more challenging at smaller values of $\kappa$; and with results
for just two values, some trends with variation of $\kappa$ may be
deduced. Sec.~\ref{sec:Interpretation} offers various decompositions
of the free energy density to facilitate the physical interpretation
of the flux structure phase diagram. In Sec.~\ref{sec:Conclusions}
we summarize the results and note their limitations, indicate some
directions for future theoretical work, and offer suggestions for
experiments.

\section{Computational Method\label{sec:Computational-Method}}

In order to avoid unnecessary repetition of material presented in
Ref.~\onlinecite{brandt_ginzburg-landau_2005}, we will start by
presenting only as much of it needed to make our further developments
intelligible.

The standard reduced units are employed, in which lengths are in units
of the penetration depth $\lambda$, energy densities are in units
of $\mu_{0}H_{c}^{2}$ (with $H_{c}$ the thermodynamic critical field)
and magnetic induction is in units of $\sqrt{2}\mu_{0}H_{c}$. Note
that in these units the upper critical mean induction is $\kappa$.
We consider infinite films with $-d/2<z<d/2$. The GL free energy
can be expressed in gauge-invariant form (rather than in terms of
the GL order parameter $\psi=fe^{i\varphi}$ and vector potential
$\mathbf{A}$) by writing it in terms of the square of the order parameter,
$\omega=f^{2}$, the supervelocity, \begin{equation}
\mathbf{Q}=\mathbf{A}-\kappa^{-1}\nabla\varphi\label{eq:defQ}\end{equation}
and the deviation from mean induction, \begin{equation}
\mathbf{b}=\mathbf{B}-\hat{z}\bar{B}\label{eq:bdef}\end{equation}
We let $S$ denote both the unit cell area and the unit cell itself,
depending on context; for the former, $S=\bar{B}/\Phi_{0}$ with $\Phi_{0}=2\pi/\kappa$.
The free energy per unit volume of superconductor referenced to the
normal state is \begin{equation}
F=\frac{1}{Sd}\int_{S}\mathrm{d}x\mathrm{d}y\int_{-d/2}^{d/2}\mathrm{dz}\left[-\omega+\frac{1}{2}\omega^{2}+\frac{\left|\nabla\omega\right|^{2}}{4\kappa^{2}\omega}+\omega Q^{2}+b^{2}\right]+\frac{2}{Sd}\int_{S}\mathrm{d}x\mathrm{d}y\int_{d/2}^{\infty}\mathrm{dz}[B^{2}-\bar{B}^{2}]\label{eq:FGL}\end{equation}
where the contribution of the first two terms in the first integral
is the condensation free energy $F_{\mathrm{cond}}$, that of the
next two terms is the kinetic energy of the supercurrent $F_{\mathrm{kin}}$,
that of the last term is the internal field energy $F_{\mathrm{mag}}$,
and that of the second integral is the stray field energy $F_{\mathrm{stray}}$.
In order to determine the phase diagram we will compare the minimum
$F$ for different vortex lattice structures with the same value of
$\bar{B}$ (and hence $S$). For a particular vortex lattice structure,
minimizing the GL free energy with respect to variations in the order
parameter yields the first GL equation\begin{equation}
\frac{1}{2\kappa^{2}}\left(\nabla^{2}\omega-\frac{\left|\nabla\omega\right|^{2}}{2\omega}\right)=-\omega+\omega^{2}+\omega Q^{2}\label{eq:firstGL}\end{equation}
The second GL equation is \begin{equation}
\nabla\times\mathbf{B}=-\omega\mathbf{Q}\label{eq:secondGL}\end{equation}
which is identical to Ampere's law in reduced units because the supercurrent
$\mathbf{j}$ is\begin{equation}
\mathbf{j}=-\omega\mathbf{Q}\label{eq:supercurrent}\end{equation}

A key step in Brandt's approach is to decompose the supervelocity
as\begin{equation}
\mathbf{Q}=\mathbf{Q}_{A}+\mathbf{q}\label{eq:Qdecomp}\end{equation}
where $\mathbf{Q}_{A}$ is the supervelocity of the Abrikosov solution
corresponding to the given vortex lattice, which satisfies \begin{equation}
\nabla\times\mathbf{Q}_{A}=\left[\bar{B}-\Phi_{0}\sum\delta_{2}\left(\mathbf{r}_{\bot}-\mathbf{R}_{\mathrm{vortex}}\right)\right]\hat{z}\label{eq:curlQA}\end{equation}
where $\mathbf{r}_{\bot}=(x,y)$, $\delta_{2}$ is the two-dimensional
Dirac delta function, and the sum runs over all vortices. Inside the
film, \begin{equation}
\mathbf{b}=\nabla\times\mathbf{q}\label{eq:bqrelation}\end{equation}

With these definitions and relations, the problem is to determine
$\omega$, $\mathbf{q}$ and $\mathbf{b}$ that minimize the free
energy. Brandt's Ansatz for these fields is as follows: \begin{equation}
\omega\left(\mathbf{r}\right)=\sum_{\mathbf{K}_{\bot},K_{z}}a_{\mathbf{K}_{\bot}K_{z}}\left[1-\cos\mathbf{K}_{\bot}\cdot\mathbf{r}_{\bot}\right]\cos\left(K_{z}z\right)\label{eq:BrandtomegaAnsatz}\end{equation}

\begin{equation}
\mathbf{q}\left(\mathbf{r}\right)=\sum_{\mathbf{K}_{\bot},K_{z}}b_{\mathbf{K}_{\bot}K_{z}}\frac{\hat{z}\times\mathbf{K_{\bot}}}{K_{\bot}^{2}}\sin\mathbf{K_{\bot}}\cdot\mathbf{r_{\bot}}\cos K_{z}z\label{eq:BrandtqAnsatz}\end{equation}

\begin{equation}
b_{z}\left(\mathbf{r}\right)=\sum_{\mathbf{K}_{\bot},K_{z}}b_{\mathbf{K}_{\bot}K_{z}}\cos\mathbf{K}_{\bot}\cdot\mathbf{r}_{\bot}\cos K_{z}z\label{eq:BrandtbzAnsatz}\end{equation}
 \begin{equation}
\mathbf{b_{\bot}}\left(\mathbf{r}\right)=\sum_{\mathbf{K}_{\bot},K_{z}}b_{\mathbf{K}_{\bot}K_{z}}\frac{\mathbf{K}_{\bot}K_{z}}{\left|\mathbf{K}_{\bot}\right|^{2}}\sin\mathbf{K}_{\bot}\cdot\mathbf{r}_{\bot}\sin K_{z}z\label{eq:BrandtbperpAnsatz}\end{equation}
Here $\mathbf{K}_{\bot}$ is the set of reciprocal lattice vectors
for the vortex lattice and $K_{z}=\left(2\pi/d\right)n$ with $n$
running over the whole numbers. Several features of this Ansatz are
worth noting. Only two sets of expansion coefficients, $a_{\mathbf{K}_{\bot}K_{z}}$
and $b_{\mathbf{K}_{\bot}K_{z}}$, are required because $\mathbf{b}$
and $\mathbf{q}$ are linked by \eqref{eq:bqrelation}. The periodicity
of the $\omega$ combined with the quadratic behavior of $\omega$
near the vortices suggests the form of expansion for the $\mathbf{r}_{\bot}$
dependence in \eqref{eq:BrandtomegaAnsatz}, while the boundary condition
for the order parameter at a superconductor-insulator interface makes
the cosine expansion natural for the $z$ dependence. Eq.~\eqref{eq:BrandtqAnsatz}
leads to supercurrents with, as one would anticipate, only in-plane
components, as well as with the appropriate periodicity and behavior
near vortex lines. The motivation for the $z$ dependence of the expansions
for $\mathbf{b}$ and $\mathbf{q}$ is that $\mathbf{q}$ and $b_{z}$
are even functions of $z$ while $\mathbf{b_{\bot}}$ is an odd function.

Inserting \eqref{eq:BrandtomegaAnsatz} and \eqref{eq:BrandtqAnsatz}
into the first GL equation and applying orthogonality of trigonometric
functions leads to coupled nonlinear equations for the expansion coefficients
$a_{\mathbf{K}_{\bot}K_{z}}$ and $b_{\mathbf{K}_{\bot}K_{z}}$ which
can be readily cast in the form of equations for the $a_{\mathbf{K}_{\bot}K_{z}}$
suitable for solution by iteration: see Eq.~\eqref{eq:a_iteration}
below. More equations must come from the second GL equation inside
the film, together with $\nabla\times\mathbf{b}=0$ outside the film
and the boundary conditions on the induction. The induction above
the film satisfies \begin{equation}
B_{z}=\bar{B}+\sum_{\mathbf{K}_{\bot}}b_{\mathbf{K_{\bot}}}^{\mathrm{s}}\cos\mathbf{K}_{\bot}\cdot\mathbf{r}_{\bot}e^{-K_{\perp}(z-d/2)}\label{eq:BzstrayBrandt}\end{equation}
\begin{equation}
\mathbf{B}_{\perp}=\sum_{\mathbf{K}_{\bot}}b_{\mathbf{K_{\bot}}}^{\mathrm{s}}\frac{\mathbf{K}_{\bot}}{K_{\perp}}\sin\mathbf{K}_{\bot}\cdot\mathbf{r}_{\bot}e^{-K_{\perp}(z-d/2)}\label{eq:BperpstrayBrandt}\end{equation}
and the continuity-of-$B_{z}$ boundary condition may be expressed
as\begin{equation}
b_{\mathbf{K_{\bot}}}^{\mathrm{s}}=\sum_{K_{z}}b_{\mathbf{K}_{\bot}K_{z}}\cos dK_{z}/2\label{eq:BrandtBzcontinuity}\end{equation}
It is convenient to derive the equations for the expansion coefficients
by direct minimization of the free energy (including the stray field
energy) with respect to the $b_{\mathbf{K}_{\bot}K_{z}}$, which leads
to Eqs.~(19) through (23) of Ref.~\onlinecite{brandt_ginzburg-landau_2005},
which we will not reproduce here.

In order to carry out a calculation of the expansion coefficients
it is necessary to truncate the expansion, setting $a_{\mathbf{K}_{\bot}K_{z}}$
and $b_{\mathbf{K}_{\bot}K_{z}}$ to zero for $\mathbf{K}_{\bot}K_{z}$
outside some range. It is also necessary to approximate the integrals
that appear in the iteration equations as finite sums. Those integrals
arise from applying orthogonality relations and, ideally, the coefficient
truncation and numerical integration could be done consistently, so
that the trigonometric functions retained in the expansion are orthogonal
with respect to the numerical integration. This is done naturally
for the $z$ coordinates of the integration, by making the simplest
choice of uniform spacing. In the $xy$ plane Brandt employs a rectangular
grid for integration but a circular domain for the allowed $\mathbf{K}_{\bot}$
values. Though a rectilinear domain for $\mathbf{K}_{\bot}$ would
be more consistent we have followed Brandt's choice, on the grounds
that when $K_{\perp}$ is large the expansion coefficients ought to
be small.

What is there to object to in the method described above? In brief,
Eq.~\eqref{eq:BrandtqAnsatz} (and its corollaries Eqs.~\eqref{eq:BrandtbzAnsatz}
and \eqref{eq:BrandtbperpAnsatz}) impose periodic boundary conditions
in the $z$ direction which are not physically appropriate. According
to \eqref{eq:BrandtbperpAnsatz}, as the film surface is approached
from within, $\mathbf{b_{\bot}}\left(\mathbf{r}\right)\rightarrow0$.
This leads to a discontinuity in $\mathbf{b_{\bot}}$ across the film
boundary, as can be seen from Eq.~\eqref{eq:BperpstrayBrandt}, and
that discontinuity implies a surface current which does not exist. 

\begin{figure}
\includegraphics[scale=0.8]{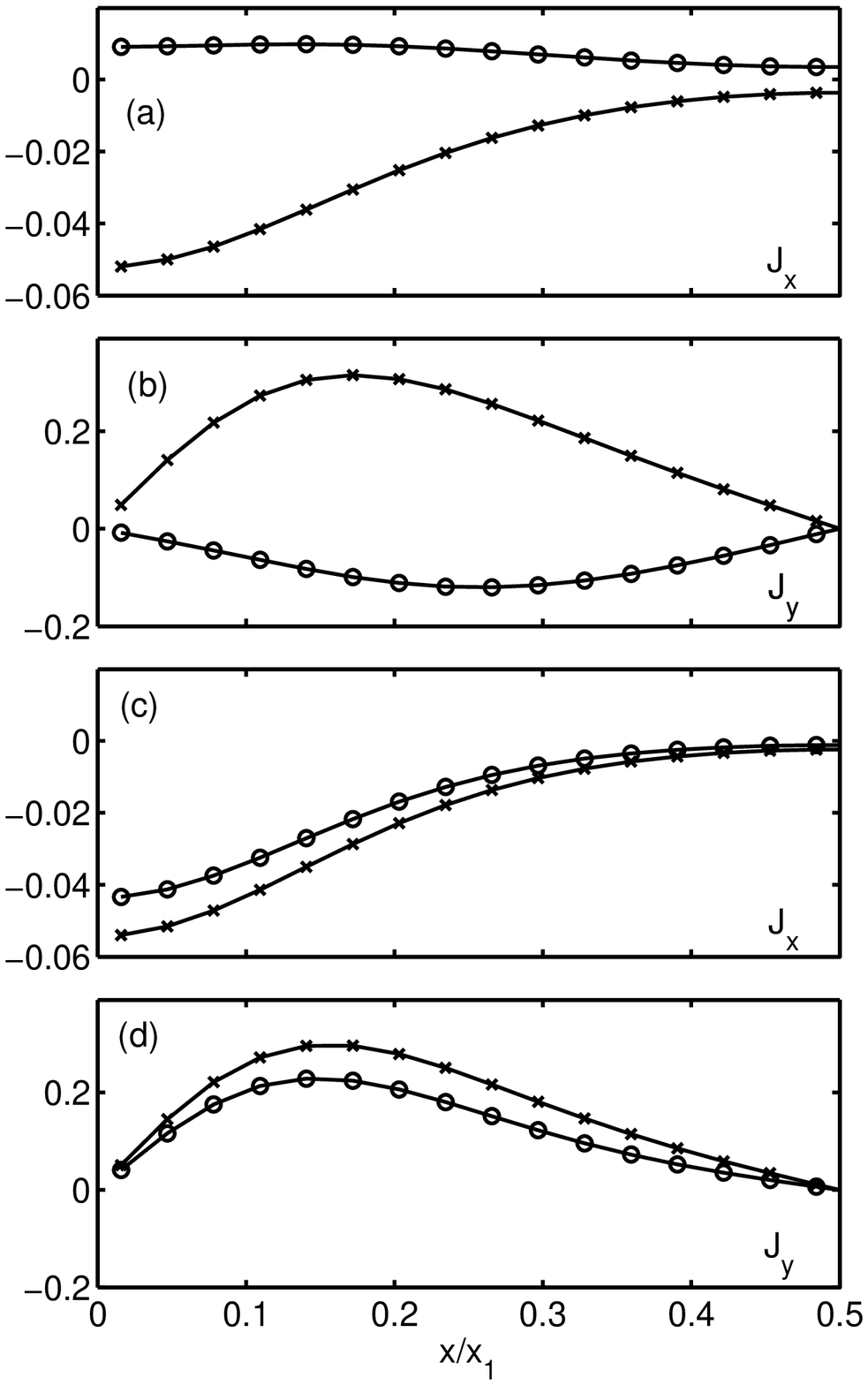}

\caption{\label{fig:Supercurrent-density-components}Supercurrent density components
$j_{x}$ and $j_{y}$ calculated from $-\omega\mathbf{Q}$ (crosses)
and $\nabla\times\mathbf{B}$ (circles) from solutions by the method
of Ref.~\onlinecite{brandt_ginzburg-landau_2005} for a system with
$\kappa=0.5$, $\bar{B}=0.4/\kappa$, $d=4.3$, and a $32\times13\times9$
grid for real-space sampling. The vortex lattice is triangular, with
one primitive translation being $x_{1}\hat{x}$. In these plots $y=0.017x_{1}$,
with $z=0.89d/2$ for (a) and (b) and $z=0$ for (c) and (d).}

\end{figure}

The consequences of this flaw in the Ansatz are surprisingly difficult
to see---no clear sign of it appears in the results presented by Brandt
in Ref.~\onlinecite{brandt_ginzburg-landau_2005}, many of which
we reproduced independently. When we implemented that method the first
suggestion of a problem came when we compared two calculations of
the supercurrent which should have given the same results, namely
$\mathbf{j}=-\omega\mathbf{Q}$ and $\mathbf{j}=\nabla\times\mathbf{B}=\nabla\times\mathbf{b}$.
An example is shown in Fig.~\ref{fig:Supercurrent-density-components},
for a system at fairly low mean induction. Note that the supercurrent
calculated according to $\nabla\times\mathbf{B}$ actually circulates
in the wrong direction for some values of $z$. A hint that the problem
was the form of the $z$ dependence in Eqs.~\eqref{eq:BrandtqAnsatz}--\eqref{eq:BrandtbperpAnsatz},
and not simply an error in our implementation as we first supposed,
was that the disagreement become more evident as the maximum value
of $K_{z}$ was increased. 

Our solution is to replace the cosine expansion for the $z$-dependence
of $\mathbf{q}$ with an expansion in terms of Legendre polynomials
of even order. Instead of Eqs.~\eqref{eq:BrandtqAnsatz}--\eqref{eq:BrandtbperpAnsatz},
take\begin{equation}
\mathbf{q}\left(\mathbf{r}\right)=\sum_{\mathbf{K_{\bot}},l}b_{\mathbf{K_{\bot}}l}\frac{\hat{z}\times\mathbf{K_{\bot}}}{K_{\bot}^{2}}\sin\mathbf{K_{\bot}}\cdot\mathbf{r_{\bot}}P_{2l}\left(2z/d\right)\label{eq:ourqansatz}\end{equation}
\begin{equation}
b_{z}\left(\mathbf{r}\right)=\sum_{\mathbf{K_{\bot}},l}b_{\mathbf{K_{\bot}}l}\cos\mathbf{K_{\bot}}\cdot\mathbf{r_{\bot}}P_{2l}\left(2z/d\right)\label{eq:ourbzansatz}\end{equation}
 \begin{equation}
\mathbf{b}_{\bot}\left(\mathbf{r}\right)=\sum_{\mathbf{K_{\bot}},l}b_{\mathbf{K_{\bot}}l}\frac{-\mathbf{K_{\bot}}}{K_{\bot}^{2}}\sin\mathbf{K_{\bot}}\cdot\mathbf{r_{\bot}}\frac{2}{d}P'_{2l}\left(2z/d\right)\label{eq:ourbperpansatz}\end{equation}
There is an additional benefit of the Legendre polynomial expansion
for the accuracy of the calculations. A numerical scheme for $z$-integration
which maintains orthogonality of the Legendre polynomials is appropriate
for the iterative calculation of the $b$ coefficients, namely, Gauss-Legendre
quadrature. The abscissas for Gauss-Legendre quadrature are at zeros
of $P_{n}$ (where $n$ is larger than the highest order used in the
Ansatz), and these zeros are more numerous near the film surfaces
where the most rapid changes occur for $\mathbf{b}$ and $\mathbf{q}$.

We now present the full scheme for generating solutions to the GL
equations for films. We use $\langle\cdots\rangle_{\mathrm{U}}$ to
denote the volume average over a unit cell by numerical quadrature
in which the $z$ abscissas are uniformly spaced, while $\langle\cdots\rangle_{\mathrm{G}}$
is the same, except it employs Gauss-Legendre quadrature for the $z$
coordinate. Angle brackets without a subscript refers to an analytic
expression for the volume average over the unit cell. Before beginning
the iterative calculations a set of initial $a_{\mathbf{K}_{\bot}K_{z}}$
and $b_{\mathbf{K}_{\perp}l}$ coefficients must be chosen; we will
discuss that choice following the iteration scheme.

For the order parameter coefficients we use Brandt's iteration scheme,
without modification, but for completeness we include it here. Defining
\begin{equation}
g=|\nabla\omega|^{2}/4\kappa^{2}\omega\label{eq:gdef}\end{equation}
the first GL equation leads to the iteration\begin{equation}
a_{\mathbf{K}_{\bot}K_{z}}:=\frac{4\langle(\omega^{2}-2\omega+\omega Q^{2}+g)\cos\mathbf{K_{\bot}\cdot r_{\bot}}\cos K_{z}z\rangle_{\mathrm{U}}}{(\delta_{K_{z},0}+1)((K_{\perp}^{2}+K_{z}^{2})/2\kappa^{2}+1)}\label{eq:a_iteration}\end{equation}
This is always followed by an iteration to minimize $F$ by multiplying
all the $a_{\mathbf{K}_{\bot}K_{z}}$ by the same factor,\begin{equation}
a_{\mathbf{K}_{\bot}K_{z}}:=a_{\mathbf{K}_{\bot}K_{z}}\langle\omega-g-\omega Q^{2}\rangle_{\mathrm{U}}/\langle\omega^{2}\rangle_{\mathrm{U}}\label{eq:a_normalization}\end{equation}
This step was introduced by Brandt in solving the GL equations in
bulk superconductors; if omitted, the calculations generally do not
converge.

Next comes the iteration for the $b_{\mathbf{K}_{\perp}l}$ . Our
modification of the expansions for $\mathbf{b}$ and $\mathbf{q}$
require corresponding changes to the iteration scheme compared to
Ref.~\onlinecite{brandt_ginzburg-landau_2005}. It is convenient
to construct some auxiliary quantities such as the stray-field expansion
coefficients \begin{equation}
b_{\mathbf{K_{\bot}}}^{\mathrm{s}}=\sum_{l}b_{\mathbf{K_{\bot}}l}\end{equation}
(compare Ref.~\onlinecite{brandt_ginzburg-landau_2005} Eqs.~(10)
and (21)); a quantity that arises from $\partial\langle\omega Q^{2}\rangle/\partial b_{\mathbf{K_{\bot}}l}$, 

\begin{equation}
D_{\mathbf{K_{\bot}}l}=\left\langle \omega\left[Q_{y}K_{x}-Q_{x}K_{y}\right]\sin\mathbf{K_{\bot}\cdot r_{\bot}}P_{2l}\left(2z/d\right)\right\rangle _{\mathrm{G}}\end{equation}
(compare Ref.~\onlinecite{brandt_ginzburg-landau_2005} Eqs.~(20)
and (22)); and %
{} \begin{equation}
S_{\mathbf{K_{\bot}}l}=\sum_{l'=0}^{l}b_{\mathbf{K_{\bot}}l'}2l'\left(2l'+1\right)+\sum_{l'=l+1}b_{\mathbf{K_{\bot}}l'}2l\left(2l+1\right)\label{eq:Scoeffdef}\end{equation}
which appears in\begin{equation}
\partial\langle b^{2}\rangle/\partial b_{\mathbf{K_{\bot}}l}=2S_{\mathbf{K_{\bot}}l}/(dK_{\perp})^{2}+b_{\mathbf{K_{\bot}}l}/(4l+1)\end{equation}
These last two expressions are rather more complicated than the corresponding
Eq.~(19) of Ref.~\onlinecite{brandt_ginzburg-landau_2005} because,
unlike sines and cosines, the $P_{2l}$ and $P'_{2l}$ are not mutually
orthogonal. The second sum in Eq.~\eqref{eq:Scoeffdef} is finite
on account of the truncation of the expansion.

With these definitions the revised iteration scheme is\begin{equation}
b_{\mathbf{K_{\bot}}l}:=\frac{-2S_{\mathbf{K_{\bot}}l}-2D_{\mathbf{K_{\bot}}l}-2K_{\bot}b_{\mathbf{K_{\bot}}}^{\mathrm{s}}/d+c\left\langle \omega\right\rangle b_{\mathbf{K_{\bot}}l}}{K_{\perp}^{2}/(4l+1)+c\left\langle \omega\right\rangle }\label{eq:b_iteration}\end{equation}
where the constant $c$ %
{}and the order parameter mean $\left\langle \omega\right\rangle =\sum_{\mathbf{K_{\bot}}}a_{\mathbf{K_{\bot}}0}$
are included to stabilize the iterations (compare Ref.~\onlinecite{brandt_ginzburg-landau_2005}
Eq.~(23)). 

The algorithm is started with an initial guess for the $a_{\mathbf{K}_{\bot}K_{z}}$
and $b_{\mathbf{K_{\bot}}l}$ coefficients. Convergence to the physical
solutions is not guaranteed, and in fact it is essential to have good
initial values. We have used bulk solutions\cite{brandt_precision_1997}
as initial values for $a_{\mathbf{K}_{\bot}0}$ and $b_{\mathbf{K_{\bot}}0}$,
with the other coefficients initially zero. One repeatedly cycles
through Eqs.~\eqref{eq:a_iteration}, \eqref{eq:a_normalization},
and \eqref{eq:b_iteration} until $F$ has converged to an absolute
tolerance of $1\times10^{-10}$ or better, which typically requires
about 200 iterations. This is slower convergence than is achieved
with the cosine Ansatz for the $z$ dependence for the supervelocity.
A possibly related matter is that we have not found a suitable expression
for the {}``mixing parameter'' $c$ that works well---large enough
to maintain stability of the iteration scheme, small enough to allow
for reasonably quick convergence---over the entire range of parameters
that we have studied. What we do instead is to adjust $c$ during
the iteration cycle by monitoring the evolution of $F_{\mathrm{mag}}$
and $F_{\mathrm{stray}}$. We have found when either of those field
energies increases excessively it is a sign that an instability is
developing. A scheme that works reliably is that when either $F_{\mathrm{mag}}$
and $F_{\mathrm{stray}}$ increases by more than 50\% following \eqref{eq:b_iteration}
then $c$ is multiplied by 10 and the $b$-iteration is re-run; independently,
every 30 iterations $c$ is divided by 2. 

Although our calculations do not converge as rapidly as those reported
in Ref.~\onlinecite{brandt_ginzburg-landau_2005} they always lead
to solutions with lower free energies, typically by 0.5\% or less
(with the same number of coefficients included in both calculations).
These small differences are enough to produce noticeable changes in
the phase boundaries. Our calculations also have the appealing feature
that increasing the $l$ cutoff for the $b_{\mathbf{K_{\bot}}l}$
always gives an improved solution; the same is not true of increasing
the $K_{z}$ cutoff for the $b_{\mathbf{K_{\bot}}K_{z}}$. Repeating
the calculations presented in Fig.~\ref{fig:Supercurrent-density-components}
yields supercurrent densities from $-\omega\mathbf{Q}$ and $\nabla\times\mathbf{B}$
which are nearly coincident, and which are close to the $-\omega\mathbf{Q}$
values displayed in that figure.

All of the results presented in the following sections are for calculations
at $\kappa=0.5$ and 0.25; even with just those two values for the
GL parameter some trends with decreasing $\kappa$ are evident. Calculations
at small $\kappa$ are considerably more challenging: we have not
yet been able to obtained converged solutions at $\kappa=0.1$.

\section{Phase Diagrams and Physical Properties\label{sec:Results}}

We have carried out a series of calculations at various values of
$\bar{B}$ and $d$, and for several kinds of vortex lattices including
triangular, square, rectangular (at various aspect ratios) and two
classes of oblique lattices which we will refer to as rhombohedral
(which interpolate between triangular and square at fixed unit cell
area, maintaining equality of the primitive vector lengths) and sheared-triangular
(which interpolate between triangular and rectangular at fixed unit
cell area, keeping one primitive vector fixed). The common feature
of the structures considered is that they have one vortex per unit
cell, and consequently the coefficients in the expansion of the order
parameter \eqref{eq:BrandtomegaAnsatz} are known for a bulk system
just below the upper critical field.\cite{brandt_treatment_1969}
The vortex structure with lowest free energy turns out to be either
triangular, square, or rectangular. 

\begin{figure}
\includegraphics{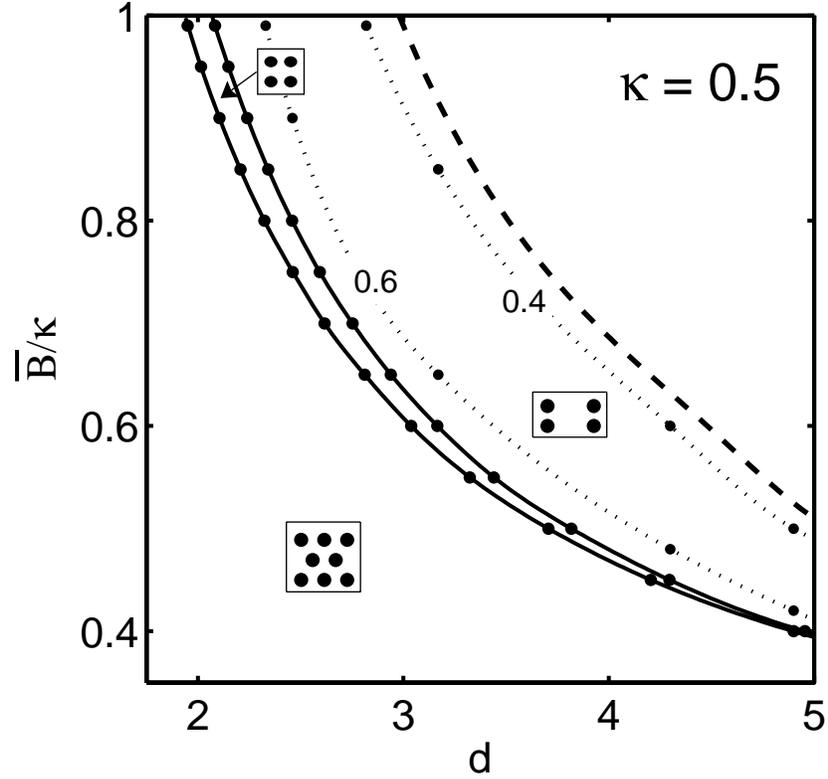}

\caption{\label{fig:PhaseDiagramk0.5}Vortex lattice phase diagram for $\kappa=0.5$.
The triangular-square transition is discontinuous while the square-rectangular
transition is continuous. Inside the rectangular phase, the dotted
lines labeled 0.6 and 0.4 are contours of constant aspect ratio. The
dashed line corresponds to the aspect ratio of 0.36, which is the
smallest aspect ratio for which a rectangular lattice is stable at
the upper critical field ($\bar{B}/\kappa=1$), following Callaway.\cite{callaway_magnetic_1992}}

\end{figure}

\begin{figure}
\includegraphics{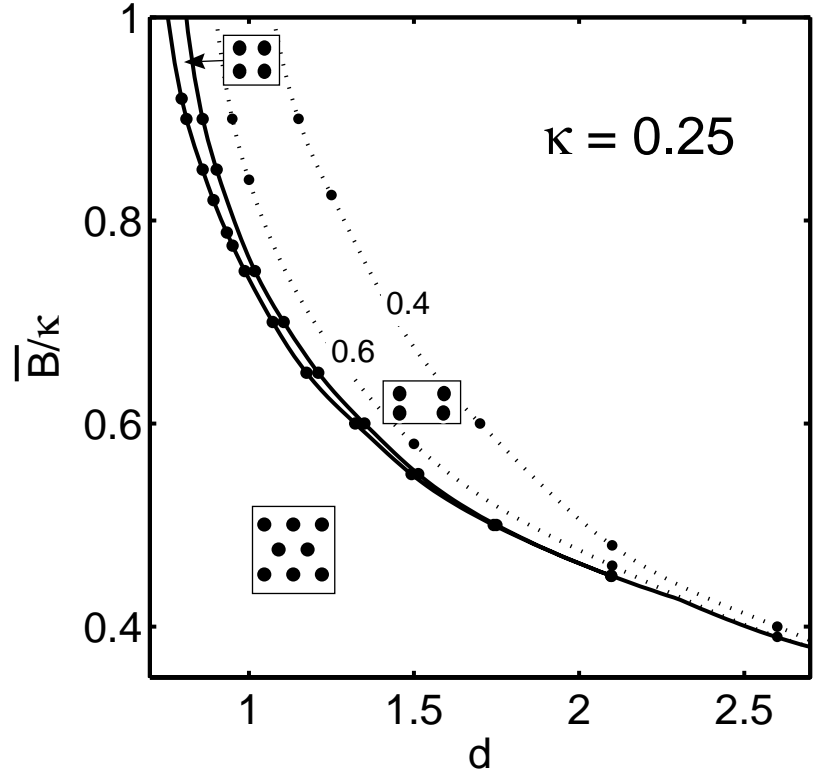}

\caption{\label{fig:PhaseDiagramk0.25}Same as Fig.~\ref{fig:PhaseDiagramk0.5},
but for $\kappa=0.25$. Note the critical endpoint for the square-rectangular
transition at $d\approx2.1$.}

\end{figure}

Figs.~\ref{fig:PhaseDiagramk0.5} and \ref{fig:PhaseDiagramk0.25}
show the resulting phase diagrams for $\kappa=0.5$ and 0.25. The
phases found at the upper critical field extend to lower fields, but
with the phase boundaries shifting to larger thicknesses as $\bar{B}$
is reduced. At sufficiently low $\bar{B}$ the interval of thickness
where the square lattice is stable is seen to vanish on the $\kappa=0.25$
phase diagram; and the same almost certainly holds for $\kappa=0.5$,
but at a lower value of $\bar{B}$ than we have considered. Contours
of constant aspect ratio within the rectangular phase are shown as
dotted lines. On the $\kappa=0.5$ phase diagram we have included
a dashed line where we speculate that the rectangular phase ends and
more complicated structures with more than one flux quantum per unit
cell begin; in drawing that line we are assuming that the aspect ratio
within the rectangular phase is constant at the boundary with the
adjacent phase. 

Within linearized GL theory $d\bar{B}^{1/2}$ is constant on every
phase boundary.\cite{lasher_mixed_1967} That is not a terrible approximation,
but the numerical results are noticeably different, with the domain
of stability of the triangular phase reduced compared to the linearized
GL theory. The critical endpoint for the square to rectangular transition
is a qualitative feature that only emerges from the full GL treatment. 

\begin{figure}
\includegraphics[scale=0.7]{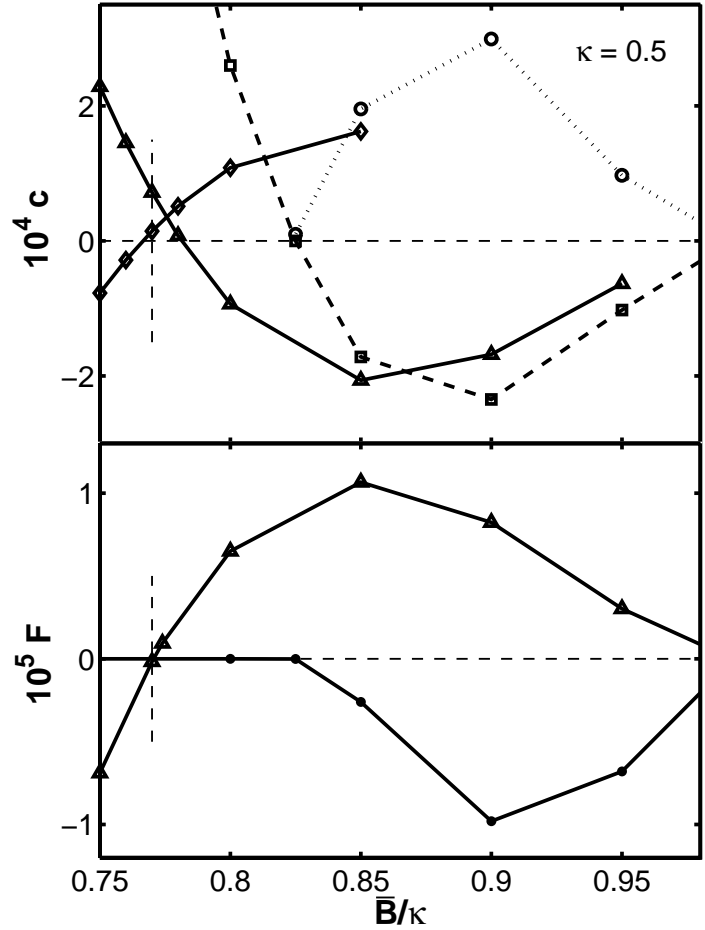}

\caption{\label{fig:SMandFk0.5}Shear moduli and free energies per unit volume
for triangular, square, and rectangular lattices, at $\kappa=0.5$
and $d=2.4$, for mean inductions around the domain of stability of
the square lattice. Free energies are referenced to value for the
square lattice; on that graph the triangular lattice values are the
triangles and the minimum-$F$ rectangular lattice values are the
circles. The vertical dashed segments in both plots indicate the transition
between triangular and square lattices, to make clear the discontinuity
in shear modulus. On the shear modulus plot, triangles are $c_{66}$
for the triangular lattice, diamonds are $c_{66}$ for the square
lattice, squares are $\frac{1}{2}(c_{11}-c_{12})$ for the square
lattice, and circles are $\frac{1}{2}((c_{11}+c_{22})/2-c_{12})$
for the minimum-$F$ rectangular lattice. Both $F$ and $c$ are in
units of $\mu_{0}H_{C}^{2}$. }

\end{figure}

\begin{figure}
\includegraphics[scale=0.7]{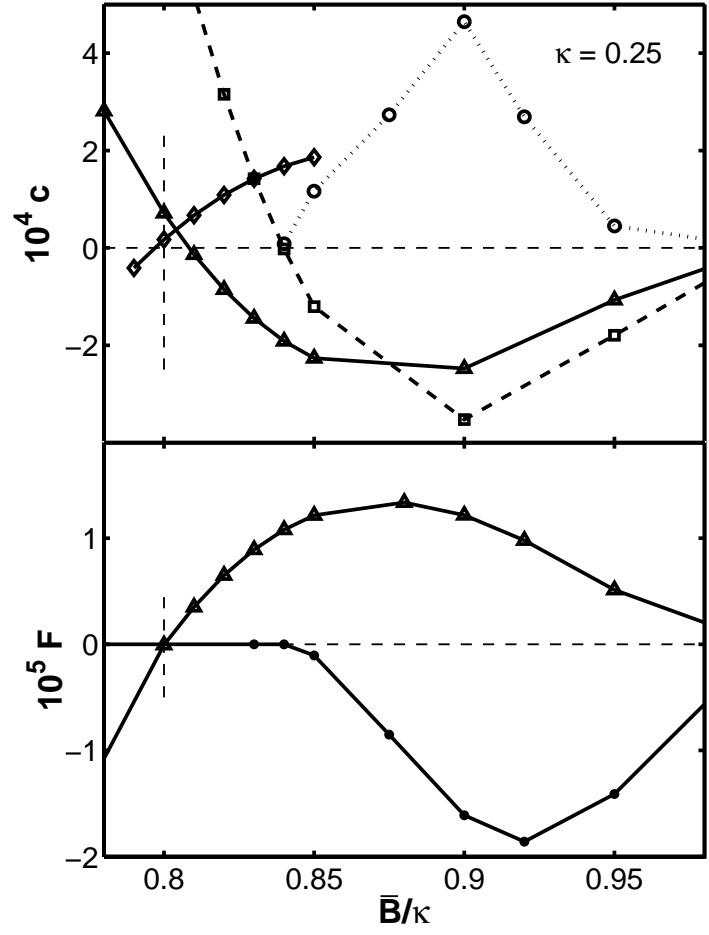}

\caption{\label{fig:SMandFk0.25}Same as Fig.~\ref{fig:SMandFk0.5}, but for
$\kappa=0.25$ and $d=0.94$.}

\end{figure}

It is of interest to look at the free energies that underlie the phase
diagrams, to see the scale of the free energy differences. In the
lower panel of Fig.~\ref{fig:SMandFk0.5}, $F$ is presented as a
function of mean induction for $\kappa=0.5$ and $d=2.4$, while Fig.~\ref{fig:SMandFk0.25}
does the same for $\kappa=0.25$ and $d=0.94$ (the latter thickness
is chosen so that the phase transitions in the two figures are at
roughly the same values of $\bar{B}/\kappa$). The rhombohedral lattice
free energies, not shown in those figures, are nearly degenerate with
the free energies of square and triangular lattices at phase transition
between them, and close to the transition their free energies almost
linearly interpolate between square and triangular lattice free energies.

Shear moduli have been evaluated for the three lattice structures
which appear on the phase diagram: see the upper panels of Figs.~\ref{fig:SMandFk0.5}
and \ref{fig:SMandFk0.25}. For triangular lattices the only shear
modulus is $c_{66}=\frac{1}{2}(c_{11}-c_{12})$. For square lattices
there are two distinct types of shear, with moduli $c_{66}$ and $\frac{1}{2}(c_{11}-c_{12})$:
the former preserves equality of primitive lattice vector length,
while the latter preserves orthogonality of primitive lattice vectors.
We present both on the figures because the latter vanishes at the
continuous square-rectangular transition and the former is anomalously
small at the discontinuous triangular-square transition. For the rectangular
lattices we considered only the shear mode which preserves orthogonality
of primitive lattice vectors; the corresponding modulus is $\frac{1}{2}((c_{11}+c_{22})/2-c_{12})$.
In every case the shear modulus is calculated by evaluating the energy
difference between the reference lattice structure and a slightly
sheared lattice. One can see in the figures the small domains of metastability
for the triangular and square lattice phases. It is also apparent
that the vortex lattices at these values of $\kappa$ and $d$ are
anomalously soft for a wide range of mean inductions.

\section{Free Energy Decompositions\label{sec:Interpretation}}

The preceding section presented the main physical results of the calculations,
but further insight might be gained by comparing not just $F$ for
different lattice structures but also various {}``components'' of
the free energy.

One decomposition is into the condensation, kinetic, and magnetic
terms described following Eq.~\eqref{eq:FGL}. Let us first consider
$\kappa=0.5$, $\bar{B}/\kappa=0.825$, and $d=2.0$, which is in
the triangular phase but not far from the square phase. For a bulk
system at the same GL parameter and mean induction, the square lattice
has lower free energy density than the triangular lattice. Why is
the relative stability reversed?
In Table~\ref{tab:Film-minus-bulk} we present the differences in
free energy density components between the film and the bulk system
for both triangular and square vortex lattices. The signs of all those
differences may be understood as a consequence of suppression of the
order parameter in the film compared to the bulk. However, the exchange
of stability is a more subtle matter, since that depends on the difference
(triangular minus square lattice values) of those free energy density
differences. Alternatively, we can compare the triangular and square
lattice free energy density components for films of different thickness,
as presented in Table~\ref{tab:tri-minus-sq}. It is then evident
that with increasing thickness, the transition to the square vortex
lattice is favored only by the condensation term.

\begin{figure}
\includegraphics{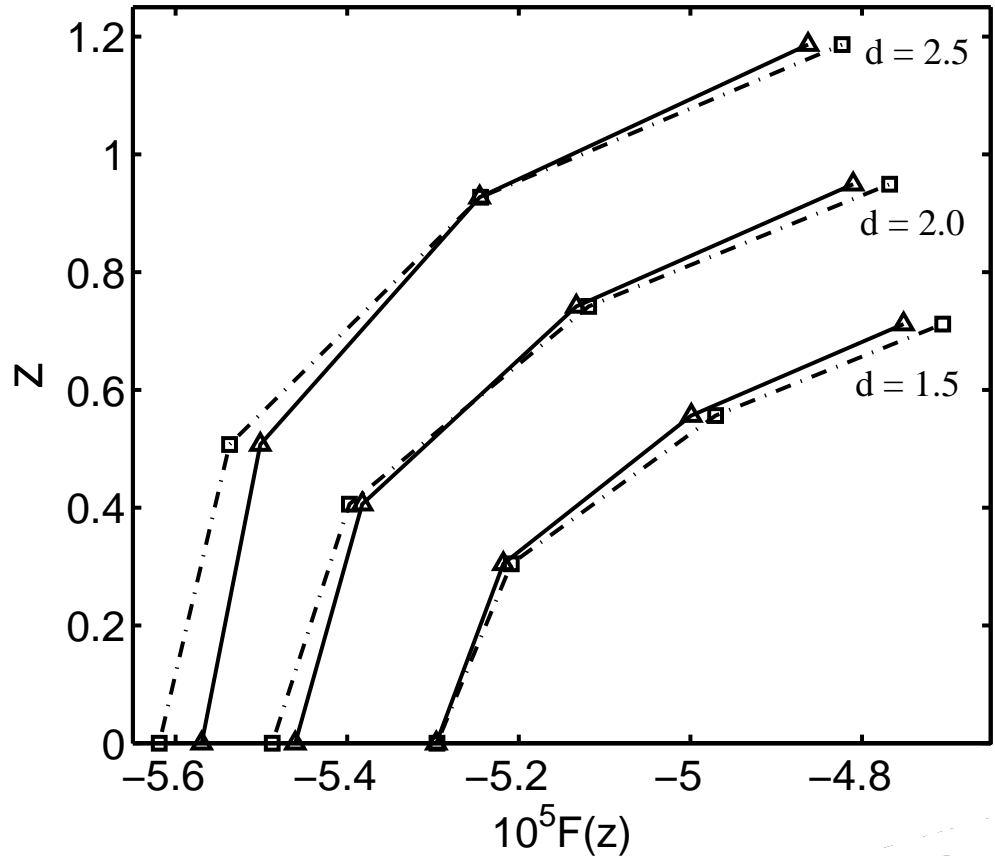}\caption{\label{fig:Fzdecom}Free energy density dependence on $z$ for films
of various thickness with $\kappa=0.5$ and $\bar{B}=0.99/\kappa$,
for square and triangular vortex lattices (indicated by the squares
and triangles, lines are guides for the eye).}

\end{figure}

We can also examine the $z$-dependence of the free energy density,
integrating in Eq.~\eqref{eq:FGL} only over $x$ and $y$ and dividing
only by $S$ to define $F(z)$. ($F_{\mathrm{stray}}$ is taken as
a $z$-independent contribution to $F(z)$.) Figure~\ref{fig:Fzdecom}
compares square and triangular vortex lattices for $\kappa=0.5$ just
below the upper critical field for $d=1.5$, 2.0, and 2.5. The triangular
lattice has lower total free energy only for $d=1.5$. However, in
every case $F(z)$ is lower for the triangular lattice when $z\approx d/2$,
and, with decreasing $z$, $F(z)$ decreases more rapidly for the
square lattice than for the triangular lattice. Figure~\ref{fig:Fzdecom}
is thus consistent with the interior of the film being more bulk-like
than the surface; and in fact $F(0)$ approaches $F$ for a bulk system
with as $d$ increases. In terms of the free energy components, the
condensation term is nearly independent of $z$, as is $\omega$ (as
pointed out by Brandt for films of type-II superconductors\cite{brandt_ginzburg-landau_2005}).
The kinetic term is responsible for the $z$-dependence seen in Figure~\ref{fig:Fzdecom},
since the magnetic term is smaller at the surface, where the field
lines spread out, than in the center of the film.

\section{Conclusions\label{sec:Conclusions}}

We have improved Brandt's method\cite{brandt_ginzburg-landau_2005}
for solving the GL equations for thin-film superconductors in perpendicular
magnetic fields, and applied it to a series of calculations for various
vortex lattice structures with one vortex per primitive cell in type-I
superconductor films of intermediate thickness ($d\sim\lambda$).
The phase diagrams presented in Sec.~\eqref{sec:Results} are the
first step beyond the linearized theory towards the development of
an accurate equilibrium flux structure phase diagram for films of
type-I GL superconductors. The results suggest that non-triangular
flux lattice structures (square and rectangular) may arise at mean
inductions well below the upper critical value. For future work we
wish to carry out similar calculations for flux structures with more
that one vortex per primitive cell, as well as lattices of multiply-quantized
vortices and various intermediate state models; these will require
different expansions for the in-plane variation of $\omega$, $\mathbf{q}$
and $\mathbf{b}$ but Legendre function expansions for the $z$ dependence
should still be applicable.

The anomalous softness of the vortex lattice in and near the domain
of stability for the square vortex lattice offers hope that some features
of the theoretical phase diagram might be observed in critical current
measurements, in the form of a {}``peak effect''\cite{pippard_possible_1969}
well below the upper critical field. However, quantitative comparison
between the theoretical phase diagrams and experimental results will
necessarily be complicated by anisotropy and, possibly, thermal fluctuations.\cite{hove_vortex_2002} 
\begin{acknowledgments}
We have benefited from many conversations with Prof. Stuart Field. 
\end{acknowledgments}
\bibliographystyle{apsrev}

\pagebreak{}%
\pagebreak{}%
\pagebreak{}%
\pagebreak{}%
\pagebreak{}%
\pagebreak{}%

\pagebreak{}%
\begin{table}
\caption{\label{tab:Film-minus-bulk}Film minus bulk free energy density terms
for $\kappa=0.5$, $\bar{B}/\kappa=0.825$, and $d=2.0$.}
\begin{tabular}{|c|D{.}{.}{1}|D{.}{.}{1}|}
\hline 
 & \mathrm{Triangular} & \mathrm{Square}\tabularnewline
\hline
\hline 
$10^{4}\,\Delta F_{\mathrm{cond}}$ & 161. & 180.\tabularnewline
\hline 
$10^{4}\,\Delta F_{\mathrm{kin}}$ & -116. & -129.\tabularnewline
\hline 
$10^{4}(\Delta F_{\mathrm{mag}}+\Delta F_{\mathrm{stray}})$ & -25.5 & -29.0\tabularnewline
\hline
\end{tabular}
\end{table}

\begin{table}
\caption{\label{tab:tri-minus-sq}Differences in free energy density terms
(triangular lattice minus square lattice) at $\kappa=0.5$ and $\bar{B}/\kappa=0.825$
for several values of $d$. Note that $\Delta F<0$ for $d=2.0$ but
is positive at the other thicknesses.}
\begin{tabular}{|c|D{.}{.}{2}|D{.}{.}{2}|D{.}{.}{2}|}
\hline 
$d$ & 2.0 & 2.33 & 2.6\tabularnewline
\hline
\hline 
$10^{4}\,\Delta F_{\mathrm{cond}}$ & -4.44 & -2.64 & -1.37\tabularnewline
\hline 
$10^{4}\,\Delta F_{\mathrm{kin}}$ & 6.55 & 5.29 & 4.42\tabularnewline
\hline 
$10^{4}(\Delta F_{\mathrm{mag}}+\Delta F_{\mathrm{stray}})$ & -2.31 & -2.60 & -2.81\tabularnewline
\hline
\end{tabular}
\end{table}

\end{document}